\documentclass{luxenote}

\title{Probing non-perturbative QED and new physics with a
LUXE-type experiment at the ILC.}
\author{Adri\'an Irles$^1$, on behalf of the LUXE collaboration}
\affil{$^1$IFIC, Universitat de Val\`encia and CSIC, C./ Catedr\'atico Jos\'e Beltr\'an 2, E-46980 Paterna, Spain}

\DocAuthors{A.~Person \\ B.~Person \\ C.~Person}
\DocEdBoard{D.~Person \\ E.~Person }
\draftversion{0.1}

\bibliography{references}

\begin{document}

\maketitle
\LUXEDocNo{LUXE-PROC-2023-003}

\begin{abstract}
The proposed LUXE experiment (LASER Und XFEL Experiment) at DESY, Hamburg, using the 16.5 GeV electron beam from the European XFEL, aims to probe QED in the non-perturbative regime created in collisions between high-intensity laser pulses and high-energy electron or photon beams. In this strong-field regime, where the electromagnetic field of the laser is above the Schwinger limit, physical electron-positron pairs will be created from the QED vacuum, similar to Hawking radiation from black holes. LUXE intends to measure the positron production rate in an unprecedented intensity regime, in and beyond the regime expected in the beam-beam interaction of future electron-positron colliders. This setup also provides a unique opportunity to probe physics beyond the standard model by leveraging the large photon flux generated at LUXE,  probing axion-like particles (ALPs) at a reach comparable to FASER2 and NA62. In this contribution, we will give an overview of the LUXE experimental setup and its challenges and explore the sensitivity of a LUXE-type experiment using the ILC’s or another future Higgs factory’s electron beam instead of the EU.XFEL one.
\end{abstract}

% Make the review table at the bottom of the title page
\vfill
\begin{center}
\textit{Talk presented at the International Workshop on Future Linear Colliders (LCWS 2023), 15-19 May 2023. C23-05-15.3.}
\end{center} 
%\makereviewtable
\clearpage

% Short documents dont always need a Table of Content / Figures / Tables, so comment out what is not needed
%\begingroup
%\color{black}
%\tableofcontents
%\listoffigures
%\listoftables
%\endgroup
%\pagebreak

%\LuxeAffil{a}{University of A}
%\LuxeAffil{b}{Institute B}
%\LuxeAuthorContributor{A.~Person}{a}{Contribution 1}
%\LuxeAuthorContributor{B.~Person}{b}{Contribution 2}
%\PrintLuxeContribute{0.30}

%\pagebreak

%Line numbers. For final version, please comment this out.
%\linenumbers

\section{LUXE and the study of strong-field QED}

The LUXE (Laser Und XFEL Experiment) aims to study Quantum Electrodynamics, QED, in uncharted regimes with very strong fields above the critical QED field strength, also known as the \textit{Schwinger limit} (in case of an electric field $E_{\text{cr}} = m^2_ec^3/(e\hbar)\approx1.32\times10^{18}\,\text{V/m}$)\footnote{Here, $m_e$ denotes the electron mass, $c$ the speed of light in vacuum, $e$ the electron charge and $\hbar$ the reduced Planck constant.}~\cite{Schwinger:1951}.  For a recent review of strong-field QED, SFQED, processes and effects, see~\cite{Fedotov:2022ely}.
Two key parameters for LUXE and the study of SFQED are the classical non-linearity parameter or laser intensity parameter, $\xi$, and the quantum non-linearity parameter, $\chi$. The former measures the work done by the EM field over an electron Compton wavelength ($\lambdabar=\hbar/(m_{e}c)$) in units of the laser photon energy $\hbar\omega$. Whenever it is larger than unity, calculating processes at any given order in the QED coupling, $\alpha$, requires a resummation at all orders of $\xi$. The quantum non-linearity parameter characterises the field strength experienced by an electron in its rest frame and the recoil experienced by the electron-emitting a photon.
Both parameters\footnote{The field intensity parameter is defined as $\xi= \frac{m_{e}\mathcal{E}_{L}}{\omega_{L}\mathcal{E}_{crit}}$ where $\omega_{L}$ is the laser wavelength and $\mathcal{E}_{L}$ is the laser electromagnetic field strength. The quantum non-linearity parameter is defined as $\chi=\frac{e\mathcal{E}_{L}\lambdabar}{m_{e}c^{2}}$.} are dependent on the beam and laser parameters. 
%Above the Schwinger limit, with $\chi>1$ and $\xi>1$, the QED vacuum is polarised, manifesting itself in creating physical electron-positron pairs.  %LUXE aims to measure the positron production rate from high-energy photons impinging on the strong-field polarised vacuum and study the onset of non-perturbativity at weak coupling in non-linear Compton scattering.

\begin{figure}[h!!]
%\begin{minipage}{0.33\linewidth}
%\centerline{
\centering
\includegraphics[width=0.55\linewidth]{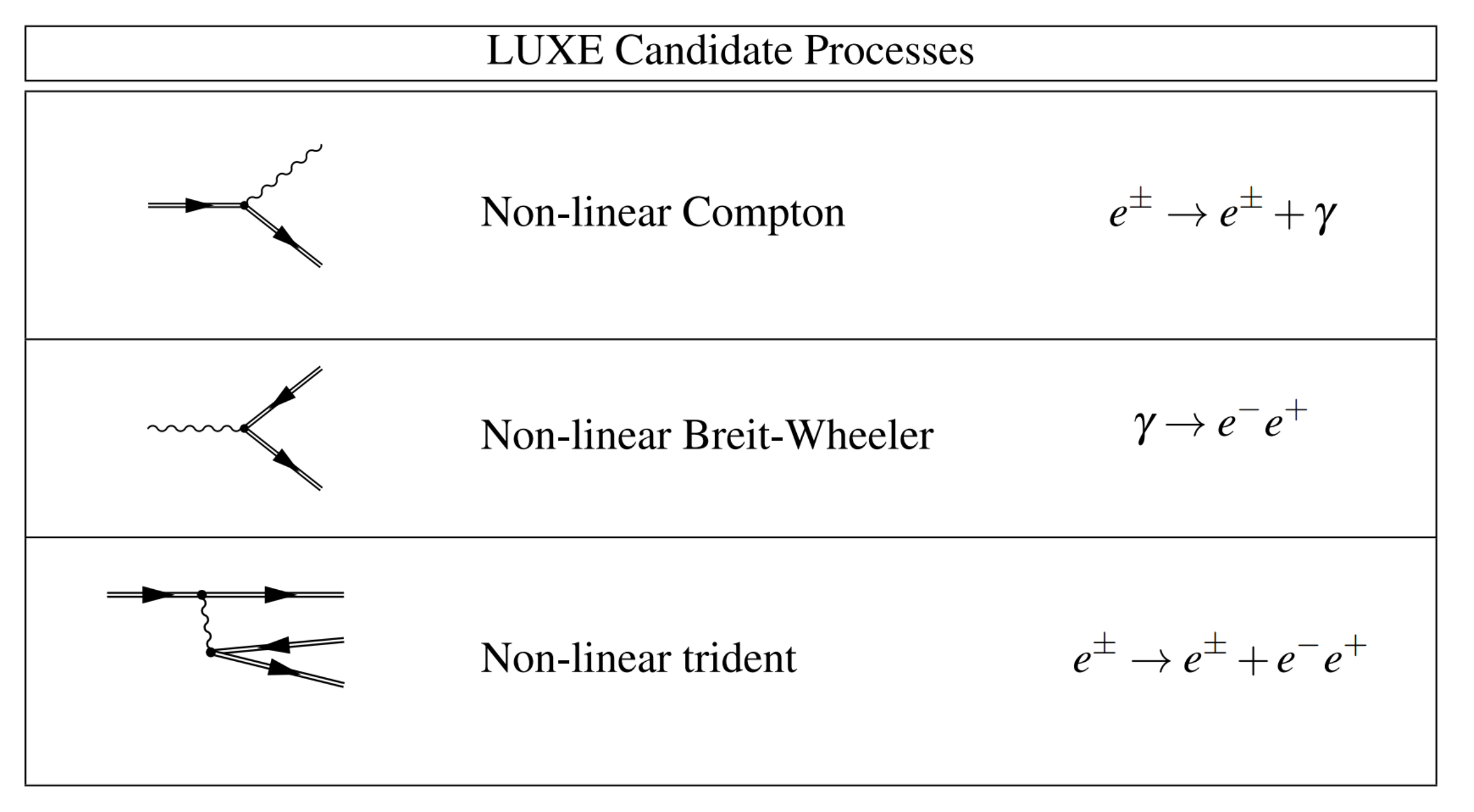} 
%\end{minipage}
\caption[]{LUXE main Strong Field QED candidate processes.}
\label{fig:sfqed}
\end{figure}

LUXE main goal is the study of the SFQED, particularly the study of non-linear Compton scattering and non-linear Breit-Wheeler and non-linear trident pair production (diagrams shown in Fig. \ref{fig:sfqed}. Non-linear Compton scattering refers to the absorption of multiple laser photons by an electron, which results in the emission of a single energetic photon. This process is examined by measuring the displacement of the Compton edge as the laser intensity parameter changes. On the other hand, the Breit-Wheeler process involves a high-energy photon absorbing multiple laser photons and producing an electron-positron pair. The scaling of this process with laser intensity is direct evidence of the transition from perturbative to non-perturbative QED and has no classical equivalent.

\begin{figure}[ht!]
%\begin{minipage}{0.33\linewidth}
%\centerline{
\centering
\includegraphics[width=0.75\linewidth]{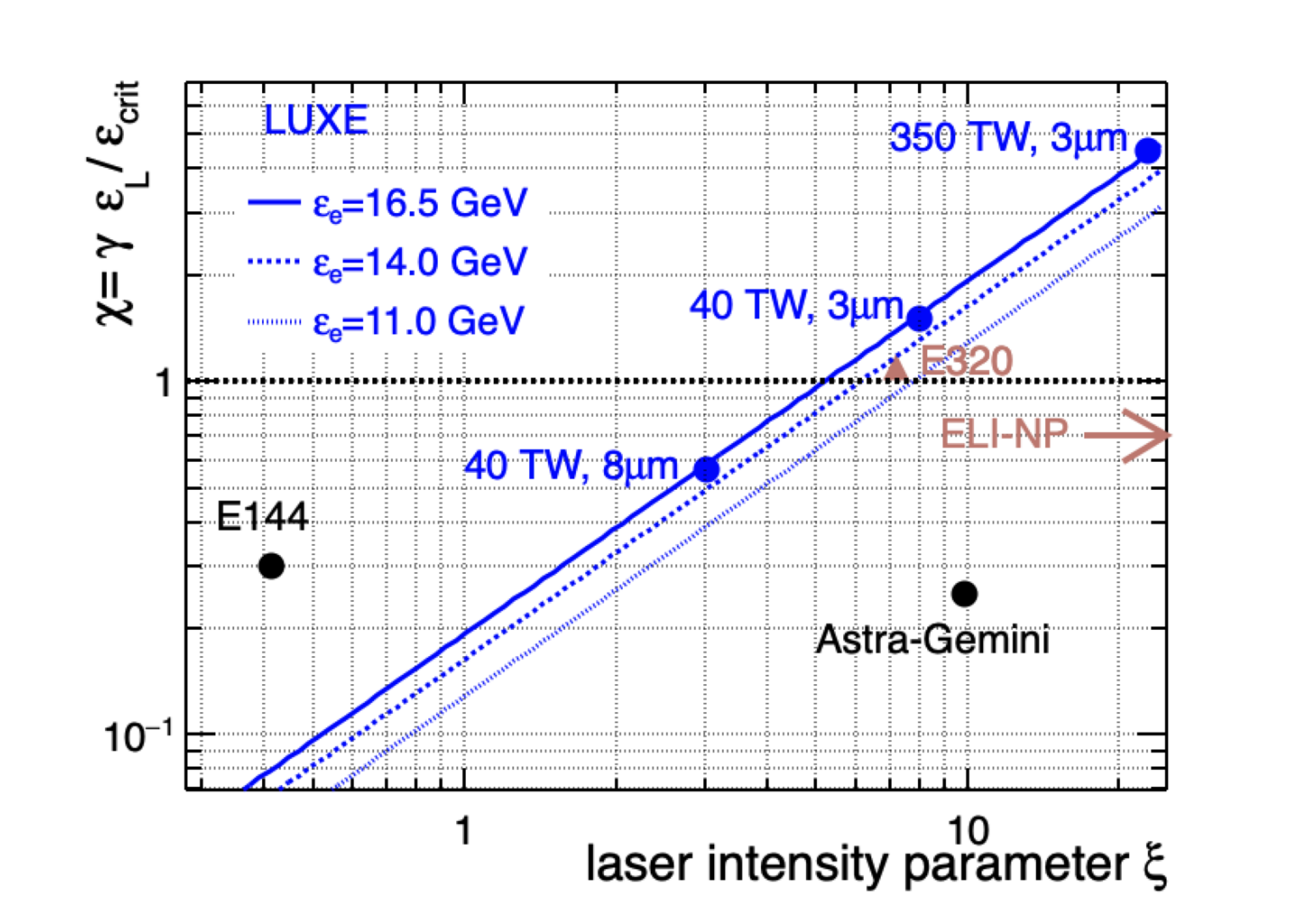} 
%\end{minipage}
\caption[]{Quantum parameter $\chi$ as a function of the intensity parameter $\xi$ for LUXE and a selection of experiments and facilities. Figure extracted from the LUXE Conceptual Design Report \cite{Abramowicz:2021zja}}.
\label{fig:luxe_reach}
\end{figure}

\subsection{Experimental setup}
SFQED fields will be reached at LUXE by creating electron-laser and photon-laser interactions with the 16.5 GeV electron beam of the European XFEL and a laser beam with a power of up to 350 TW.  
A staged approach is planned, using an upgradable laser system, which will deliver a power of $40\,\text{TW}$ in phase-0 ($\xi_\text{max}=7.9,\:\chi_\text{max}=1.5$) and subsequently will be upgraded to $350\,\text{TW}$ for phase-1 ($\xi_\text{max}=23.6,\:\chi_\text{max}=4.45$). During the data-taking in LUXE, $\xi$ and $\chi$ are varied by de-focusing and re-focusing the laser pulse at the interaction point.
The reach of LUXE at different stages of its operation is compared with the other present, past or future experiments in Fig. \ref{fig:luxe_reach}.
\\ 

\begin{figure}[h!]
\centering
\begin{tabular}{cc}
    \includegraphics[width=0.45\textwidth]{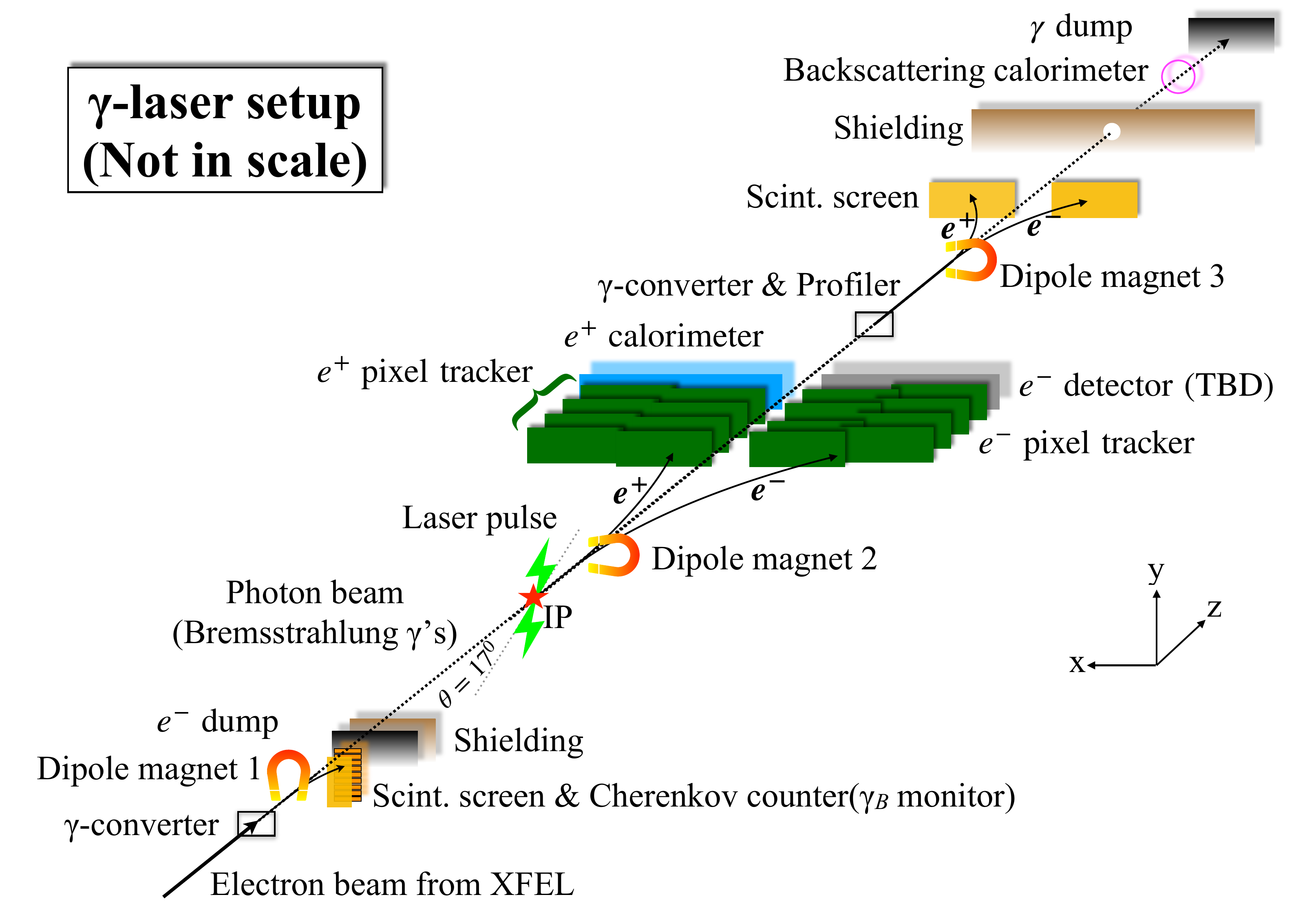} &  \includegraphics[width=0.45\textwidth]{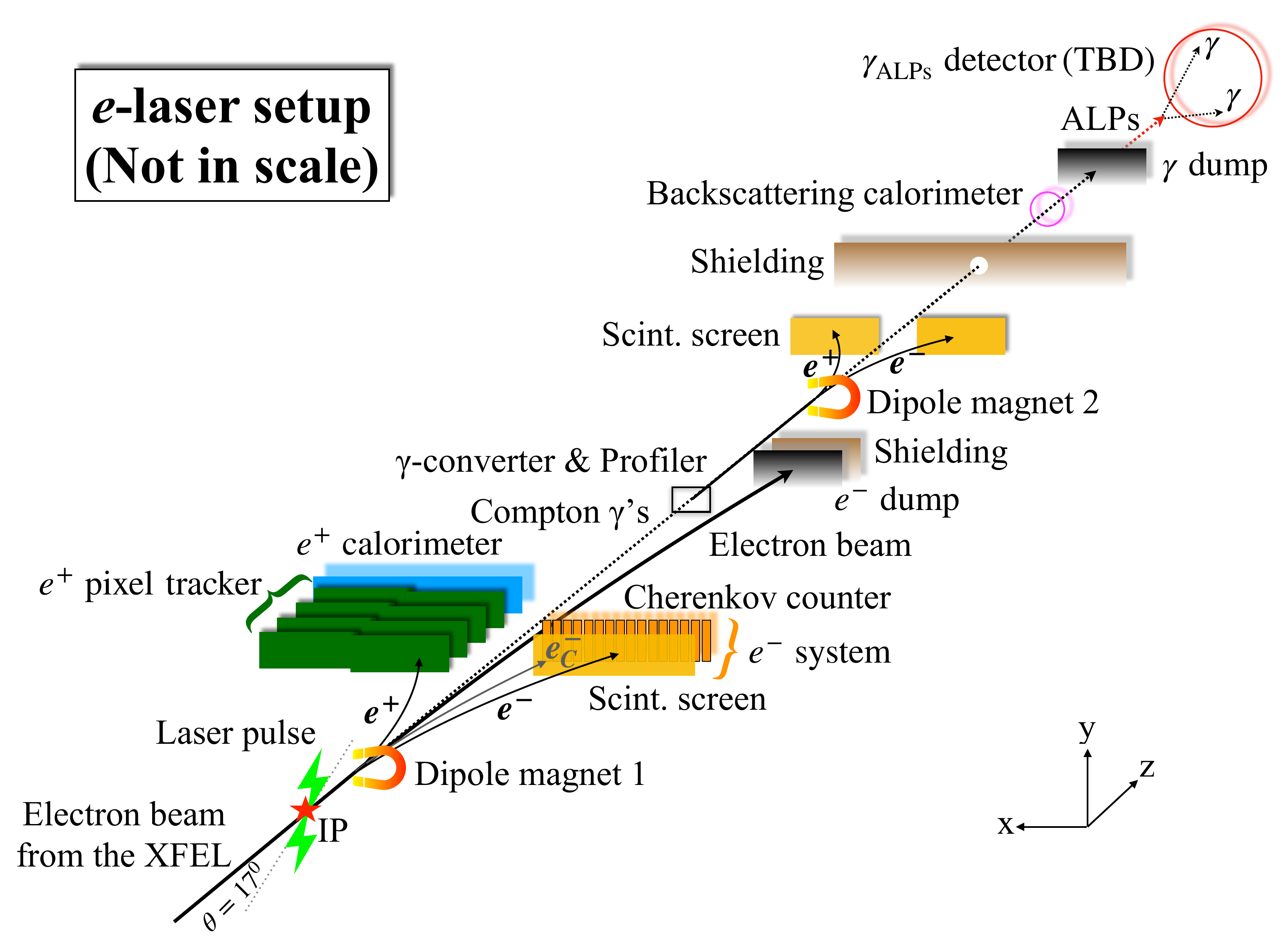} 
\end{tabular}
%\begin{minipage}{0.4\linewidth}
%\centerline{\includegraphics[width=\textwidth]{figures/detector_conceptual_layout_bppp_color_wide.pdf}
%}
%\end{minipage}
%\hfill
%\begin{minipage}{0.4\linewidth}
%\centerline{\includegraphics[width=\textwidth]{figures/detector_conceptual_layout_trident_color_wide_with_BSM.pdf}
%}
%\end{minipage}
\caption[]{Sketch of the LUXE experimental setups.}
\label{fig:operation}
\end{figure}

LUXE will run in two modes of operation: the $e$-laser mode, in which the intense laser collides directly with the electron beam, and the $\gamma$-laser mode, in which the laser interacts with secondary photons generated by the electron beam in a high-Z target upstream of the interaction point. These two operation modes are shown in Fig. \ref{fig:operation}.
In these interactions, a broad range of fluxes of electrons, positrons and photons will be produced: the expected ranges are $10^{-3}$ to $10^9$ per 1 Hz bunch crossing, depending on the laser power and focus.  %Table \ref{tab:luxerates} summarizes the particle rates for each location and run mode.\\

%\begin{table}[!h]
%\begin{center}
%\begin{tabular}{|c|c|c|} \hline
% \textbf{Location} & \textbf{$e^-$-laser particle rate} & \textbf{$\gamma$-laser particle rate} \\\hline
%  initial target monitor & $-$ &  $10^5$  \\\hline
%  post-IP $e^+$ & $10^{-4}-10^{4}$ & $10^{-4}-10^{1}$ \\\hline
%  post-IP $e^-$ & $10^{3}-10^{8}$ & $10^{-4}-10^{1}$ \\\hline
%  forward $\gamma$ & $10^{3}-10^{8}$ & $10^5$ \\\hline
%\end{tabular}
%\caption{Particle rates expected in LUXE for different locations in the experimental setup and run modes.}
%\label{tab:luxerates}
%\end{center}
%\end{table}

In addition, low-energy high radiation backgrounds will be present at LUXE. To overcome such challenges and study the SFQED in uncharted regimes with precision, LUXE foresees the use of dedicated physics-driven detectors at specific locations downstream of the interaction point.

Saludos,Providing a detailed description of these systems is out of the scope of this contribution. For a more comprehensive picture, we refer the reader to the Conceptual Design report of LUXE \cite{Abramowicz:2021zja} and the Technical Design report of LUXE, which will soon  appear.

\section{LUXE-NPOD: new physics searches with an optical dump at LUXE}
\label{sec:npod}
\begin{figure}[h!]
\centering
\begin{tabular}{c}
    \includegraphics[width=0.5\textwidth]{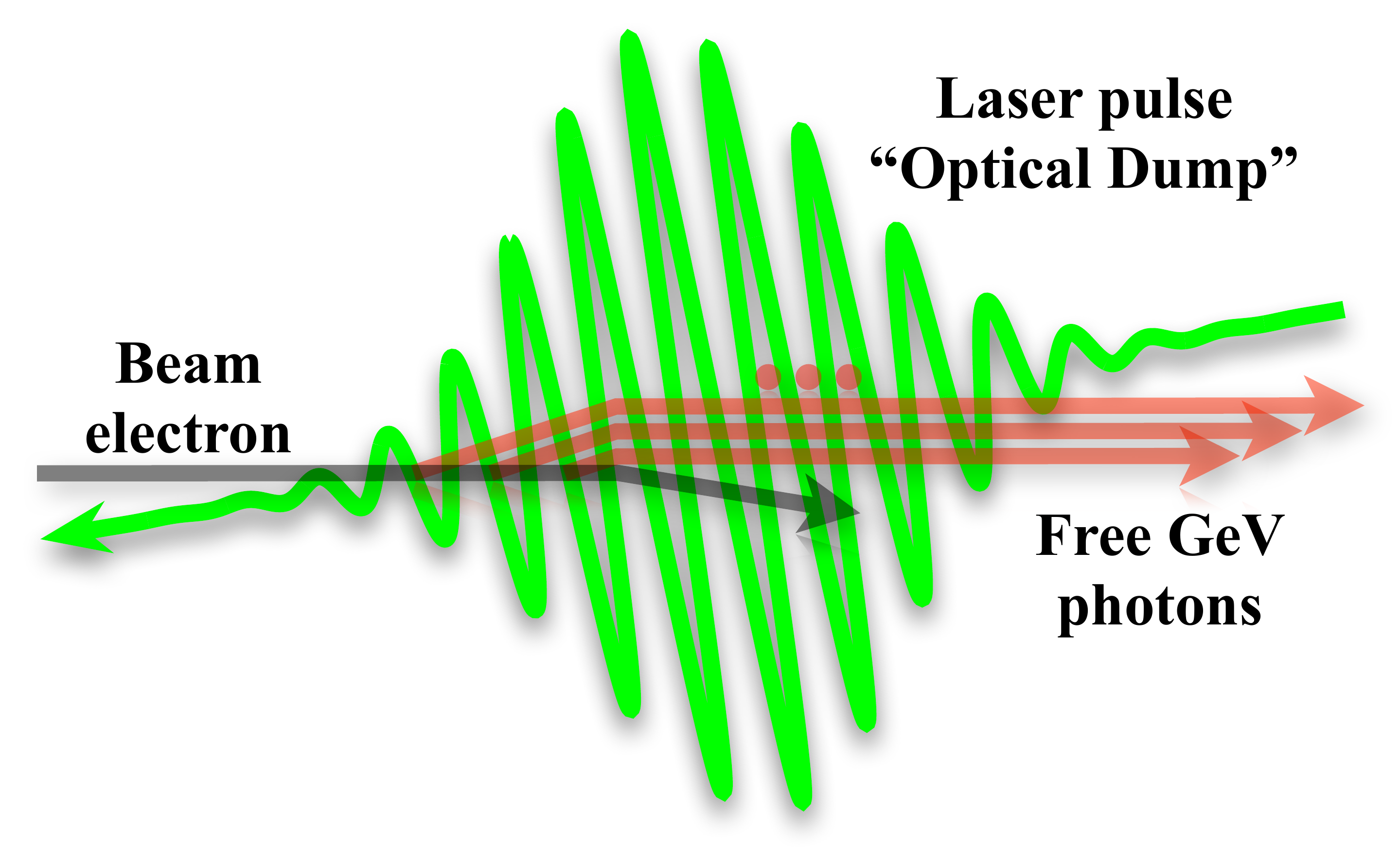} 
\end{tabular}
\caption[]{Schematic illustration of the optical dump.% The high-intensity laser pulse behaves effectively as a thick medium for the incoming electron, which may emit a large flux of hard photons which “free stream” in this optical medium and can be used to search for new physics.
}
\label{fig:OpticalMedium}
\end{figure}

The LUXE experiment will provide an intense secondary beam of hard photons through the interaction between the high-energy
electron beam and the laser pulses as optical dump. This dump behaves as thick target for the incoming electron, through non-linear Compton scattering but with negligible interaction with the photon beam. Fig. \ref{fig:OpticalMedium} shows a schematic illustration of the optical dump. Therefore, the intense and collimated hard photon beam flux can be efficiently used for specific new physics searches beyond the Standard Model, BSM. LUXE plans to use this beam to search for weakly interacting new particles that couple to photons. In particular, LUXE will provide access to direct searches of new spin-0 (scalar or pseudo-scalar)
particles with coupling to photons, as the axion-like particles (ALPs).
This proposal is denoted as LUXE-NPOD: New Physics at Optical Dump.

\begin{figure}[h!!]
\centering
\begin{tabular}{c}
    \includegraphics[width=0.6\textwidth]{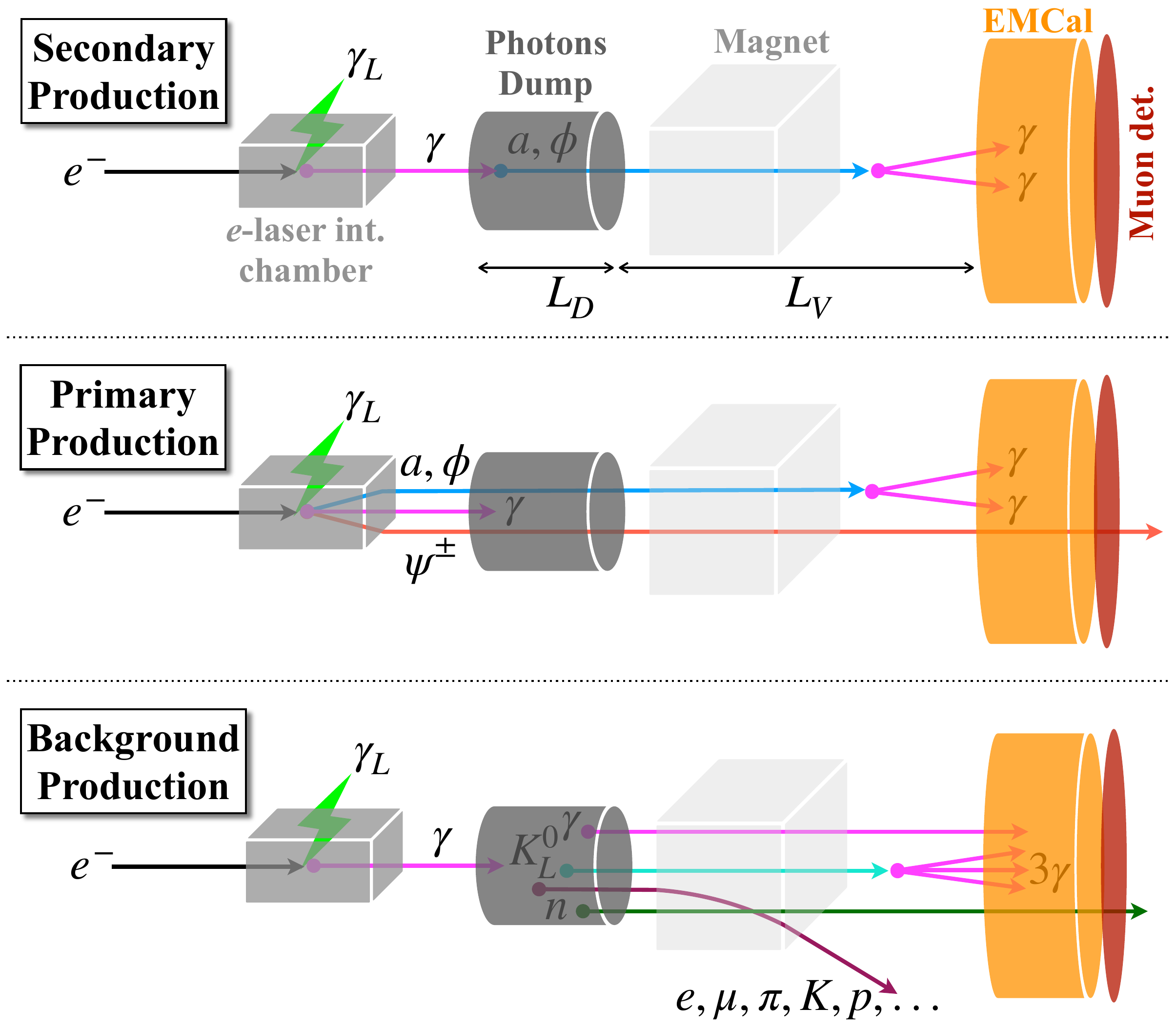} 
\end{tabular}
\caption[]{An illustration of the LUXE-NPOD concept and
the different search modes. Shown are schematics of the \textit{secondary} (top) and \textit{primary} (middle) production mechanisms realisation in the experimental setup. The relevant background topologies are also shown (bottom). The charged particles
are deflected by a magnet placed right after the interaction
chamber.}
\label{fig:BSMDet}
\end{figure}

The most relevant production mechanism for these new physics particles, NP, is the so-called \textit{secondary} NP production. In this case, the
photons produced in the electron-beam and laser pulse
collisions freely propagate through the beam pipe until they reach
a sizeable thick dump made of heavy nuclei. These NP are produced via the Primakoff production mechanism during the photon-nuclei interaction. The other production mode is the \textit{primary} NP production, in which the NP is produced directly in the electron-laser
interaction region via electron coupling. This latter mode of production is considerably suppressed compared with the  \textit{secondary} NP production, and it allows the inspection of only low mass NP $\mathcal{O}(keV)$ instead of $\mathcal{O}(GeV)$.
These production mechanisms and their kinematics are schematically represented in Fig. \ref{fig:BSMDet}.

\begin{figure}[h!!]
\centering
\begin{tabular}{c}
    \includegraphics[width=0.6\textwidth]{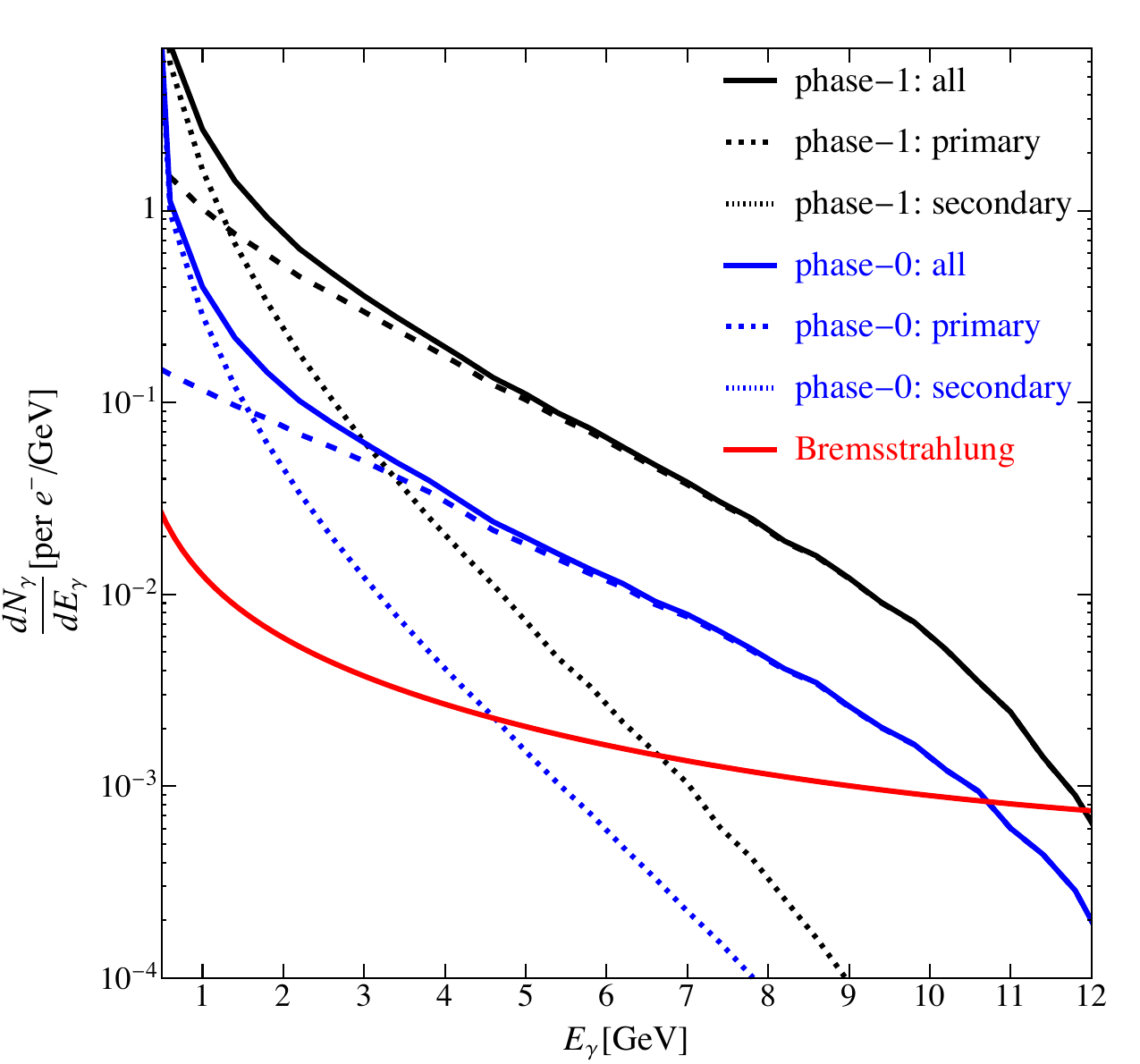} 
\end{tabular}
\caption[]{The emitted photon spectrum for phase-0 (1) in
blue (black) compared to the perturbative Bremsstrahlung
spectrum with $E_{e^{-}}= 16.5$ GeV and target length of $0.01~X_{0}$
in red.}
\label{fig:spectrum}
\end{figure}

\begin{figure}[!ht]
\centering
\begin{tabular}{c}
    \includegraphics[width=0.7\textwidth]{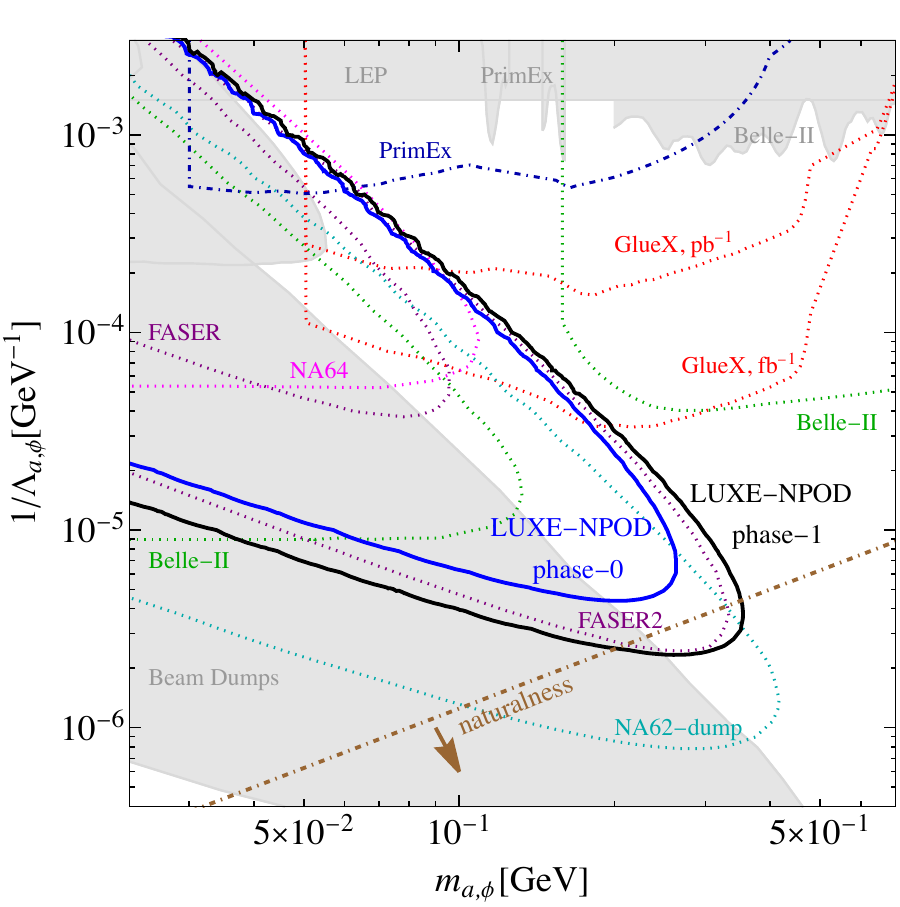} 
\end{tabular}
\caption[]{The projected reach of LUXE-NPOD phase-0 (1) in
a solid blue (black) compared to the currently existing bounds
(gray regions) and projections (dotted) on ALPs-couplings from other experiments. Details on this plot and references are to be found in \cite{Bai:2021gbm} .}
\label{fig:alps}
\end{figure}

The number of NP produced and decayed in front of the calorimeter, $N_{X}$, depends directly on effective luminosity ($\mathcal{L}_{eff}$), the Primakoff production cross section ($\sigma_{X}$), the energy spectrum of the photon beam $\frac{dN_{\gamma}}{dE_{\gamma}}$ and the acceptance $\mathcal{A}$ and geometrical characteristics of the setup and detector:
\begin{equation}
    N_{X} \simeq \mathcal{L}_{eff} \int dE_{\gamma} \frac{dN_{\gamma}}{dE_{\gamma}} \sigma_{X}(E_{\gamma}) (e^{-\frac{L_{D}}{L_{DX}}}-e^{-\frac{L_{V}+L_{D}}{L_{DX}}}) \mathcal{A}
\end{equation}
The expected $\frac{dN_{\gamma}}{dE_{\gamma}}$ distribution for LUXE operating in different modes are shown in Fig. \ref{fig:spectrum}.
The physical dump (or photon dump) at the end of the beamline is made of a block of tungsten with a length of $L_{D}=1$ m in the baseline design, and it is positioned at 13 m of the interaction point.
The NP produced in the photon-dump interaction are long-lived. Therefore, an empty distance $L_{V}=2.5$ m (baseline design) is left between the physical dump and the detector. $L_{X}$ is the propagation length of the $X$ particle. The detector should consist of an electromagnetic calorimeter with good energy and spatial resolution. It must be able to separate photon showers to reconstruct the $\gamma\gamma$ vertex and the two photons' invariant mass with precision. In addition, it should feature photon-neutron discrimination capabilities and timing of $\mathcal{O}(0.1)$ ns for the background rejection. The detailed design and technological choice are still under discussion.
The effective luminosity is directly proportional to the density and radiation length of the physical dump and to the multiplication of the number of electrons per bunch and total bunches collected. 
With these ingredients and assuming one year of data taking and a negligible amount of backgrounds, the LUXE collaboration has estimated the projected reach of LUXE-NPOD for the direct searches of new physics.
The reach of LUXE in the plane of the NP mass, $m_{X}$, versus the effective coupling, $1/\Lambda_{X}$, is shown in Fig. \ref{fig:alps} and compared with other existing or coming experiments. More details on this plot can be found in \cite{Bai:2021gbm}.
This figure shows that already in the phase-0 LUXE will probe a never reached parameter space in the mass range of 50 MeV $\lesssim m_{X} \lesssim 250$ MeV and $1/\Lambda_{X} > 4 \times 10^{-6}$ GeV$^{-1}$. For the phase-1, the parameter space is increased up to 40 MeV $\lesssim m_{X} \lesssim 350$ MeV and $1/\Lambda_{X} > 2 \times 10^{-6}$ reaching the naturalness limit for the scalar model.

\section{ LUXE-type experiment at Higgs Factories}
In this contribution, we focus on the prospects of having a LUXE-like experiment at the International Linear Collider (ILC), as studied in \cite{ILCInternationalDevelopmentTeam:2022izu}.
The ILC and Eu.XFEL beams feature similar characteristics, being both based on 1.3 GHz superconducting radio-frequency cavities producing pulsed electron and positron beams. However, the ILC and all other Higgs Factories proposals will operate electron and positron beams of the order of 120 GeV in their baseline operation modes. This is one order of magnitude larger than what is foreseen at LUXE with the EU.XFEL electron beam. With this energy, the Lorentz boost of the photon in the reference system of the accelerated electron will be more than ten times larger than at the EU.XFEL, resulting in a reach for $\chi$ of about 40 times more than LUXE. 
The tentative timeline of the ILC expects collisions in the mid-2030s. It is realistic to assume that at that time, 100 PW lasers at wavelengths of 1\textmu will be on reach. Assuming a pulses size of 1\textmu in diameter, and a pulses length of 120\textmu m, we would expect values of the SFQED quantum non-linearity parameter of $\chi\sim 250$, which is 40 times larger than at LUXE in EU.XFEL.
Moreover, the combined usage of such powerful laser and energetic electron beams will allow us to study SFQED phenomena unreachable until now as can be the creation and dynamics of coherent and incoherent $e^{+}e^{-}$ plasma.

The direct searches of new scalar or pseudo-scalar particles will also benefit from the higher beam energy and higher luminosity foreseen at Higgs Factories. 
The most critical beam parameters for a LUXE experiment at Higgs Factories are summarised in Table \ref{tab:luie}\footnote{Compiled by F. Meloni, J. List and F. Zimmerman for the  \href{https://indico.desy.de/event/33640}{\textit{$1^{st}$ ECFA Workshop on $e^{+}e^{-}$ Higgs/EW/Top Factories}}}, including prospects for linear and circular Higgs factories proposals.
The table assumes the baseline designs of four different accelerator scenarios and $10^7$ seconds of data taking per year. Of course, for ILC or FCC-ee, the beams can only be used for a LUXE-type experiment once they have been used for their main purpose (high energy $e^{+}e^{-}$ collisions). For the ILC, this study assumes the usage of all spent beams, which have a larger energy spectrum than the primary ones. For FCC-ee, two cases have been considered: the usage of the dump beams (3 times per day) or the usage of dedicated FCC-ee booster cycles for a beam dump every 10 seconds. In all cases, the same laser as in phase-0 for LUXE is assumed. The last row shows the enhancement of the signal yield for ALPs production.

\begin{table}[!h]
\begin{center}
\begin{tabular}{c|cccc} \hline
  & \multicolumn{4}{c}{\textbf{Accelerators}} \\\hline
 \textbf{Beam parameters} & \textbf{Eu.XFEL} & \textbf{ILC250} & \textbf{FCC-ee} & \textbf{FCC-ee (booster)} \\\hline
  Electron beam energy [GeV] & 16.5 & 125 & 120 & 120 \\
  Number of electrons per bunch & $1.5\times 10^{9}$ & $2\times 10^{10}$ & $1.8\times 10^{11}$ & $0.5\times 10^{10}$ \\
  Number of bunches in one year  & $10^{7}$ & $6.6\times 10^{10}$ & $1.1\times 10^{5}$ & $3.3\times 10^{8}$ \\
  Signal Yield (w.r.t. Eu.XFEL) & 1 & $8.8\times 10^{4}$ & 1.3 & $3.3\times 10^{8}$ \\
  \hline
\end{tabular}
\caption{Beam parameters for four accelerator scenarios.}
\label{tab:luie}
\end{center}
\end{table}

\begin{figure}[h!!]
\centering
\begin{tabular}{c}
    \includegraphics[width=0.6\textwidth]{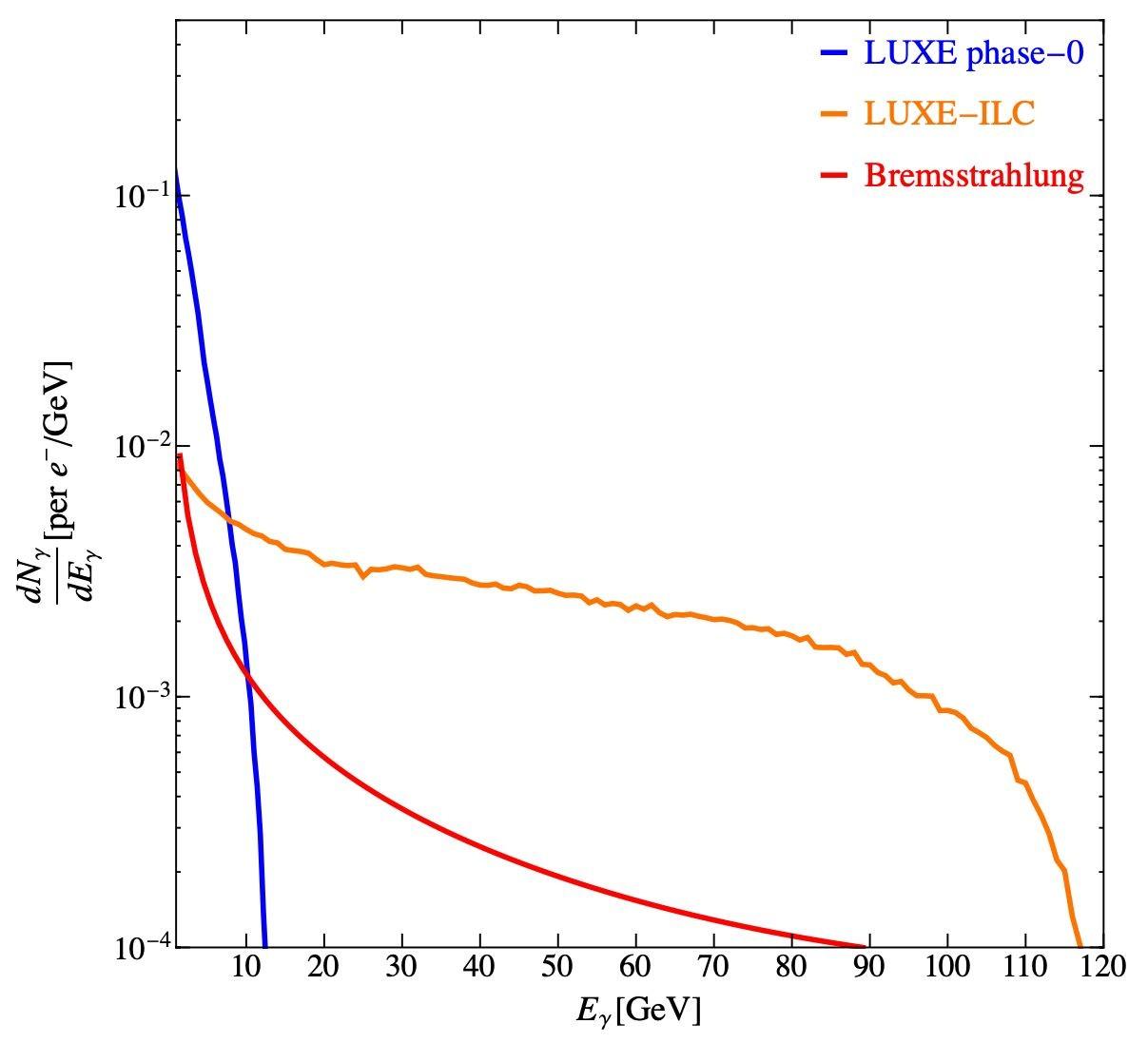} 
\end{tabular}
\caption[]{The emitted photon spectrum for a LUXE-like experiment at the ILC.
For comparison, in blue, we show the LUXE phase-0 spectrum. In red is the perturbative Bremsstrahlung
spectrum at ILC. In orange, the expected spectrum with $E_{e^{-}}= 125$ GeV.
in red.}
\label{fig:spectrum_luie}
\end{figure}

A very important aspect, granted by the higher beam energy, is the harder photon beam spectrum obtained in the laser-beam interaction, with an average $E_{\gamma}$ of 40 GeV, in contrast with the expected at LUXE. This is shown in Fig. \ref{fig:spectrum_luie} and is to be compared with Fig. \ref{fig:spectrum}.
This alone, without an upgrade of the laser setup, will allow access to the production of larger masses for the ALPs.

\begin{figure}[h!!]
\centering
\begin{tabular}{c}
    \includegraphics[width=0.7\textwidth]{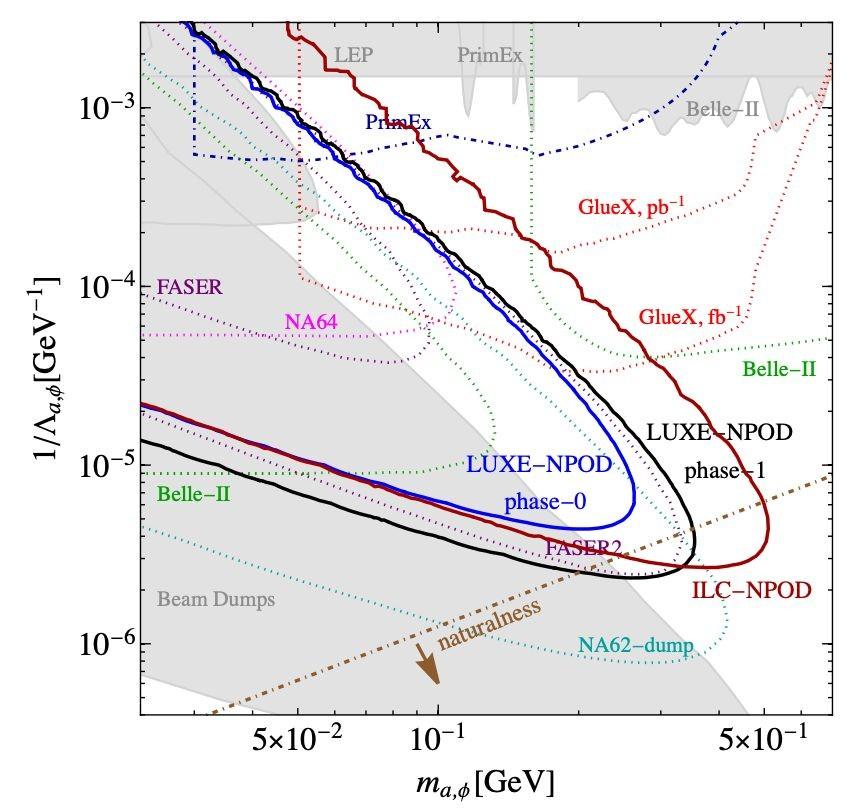} 
\end{tabular}
\caption[]{The projected reach of a LUXE-NPOD proposal operating with ILC beams. This plot is an update of Fig. \ref{fig:alps} with the addition of the study performed in \cite{ILCInternationalDevelopmentTeam:2022izu} .}
\label{fig:alps_luie}
\end{figure}

The study presented in Section \ref{sec:npod} has been extended to the ILC case. Similarly to the LUXE case, we assumed a background-free scenario but doubled the depth of the physical dump. The same laser setup as in LUXE phase-0 is assumed, and only secondary production by the Primakoff process. This leads to an enhancement of the reach in direct searches of NP, up to masses of 0.5 GeV.
The result of this study is summarised in Fig. \ref{fig:alps_luie}

Another exciting possibility is to use the photon beam generated by the ILC positron source to produce and detect the non-standard scalar or pseudo-scalar particles through the concept of "light-shining-through-the-wall". This idea and that described above are under investigation by the LUXE and ILC collaborations \cite{ILCInternationalDevelopmentTeam:2022izu}.

\section{Conclusion}
\label{sec:conclusion}
The main objective of the LUXE experiment is to explore the realm of Strong Field Quantum Electrodynamics. This will be achieved by analyzing the collisions between a high-energy photon beam or electron beam and a high-intensity optical laser. The experiment will be conducted in a continuous data-taking mode, enabling the measurement of strong-field QED processes such as non-linear Compton scattering and Breit-Wheeler pair creation with high precision. The laser system and particle detectors in LUXE are explicitly designed to cater to the physics requirements of the experiment. LUXE is poised to become the first experiment to venture into the uncharted territory of QED under conditions free from external interference, making it a significant milestone for the scientific community. Additionally, with the $\gamma$-laser setup LUXE will be the first experiment to investigate collisions between high-intensity laser and real high-energy gamma photons.

The LUXE-NPOD extension is a novel approach to detecting feebly interacting spin-0 scalar or pseudoscalar particles. This proposal has the potential to explore a challenging region of parameter space. It will use an intense GeV photon beam generated through the interactions between high-energy electrons, and a highly intense laser pulse, which can be directed towards a target dump to create new BSM states through the Primakoff process. The produced non-standard particles would have a lifetime long enough to traverse the target dump; they can decay into photon pairs that a calorimeter system can detect in an experiment that is effectively free of background noise. This experiment could help exclude new scalar or pseudoscalar states with masses ranging between 40 MeV and 350 MeV for an effective coupling of $1/\Lambda_{X} > 2 \times 10^{-6}$.

Finally, we presented first studies of the promising prospects of LUXE and LUXE-NPOD types of experiments using $\mathcal{O}(100$GeV$)$ electron beams from a future Higgs Factory as the International Linear Collider.

\section*{Acknowledgements}
\addcontentsline{toc}{section}{Acknowledgements}
We thank the DESY technical staff for continuous assistance and the DESY directorate for their strong support and the hospitality they extend to the non-DESY members of the collaboration. This work has benefited from computing services provided by the German National Analysis Facility (NAF) and the Swedish National Infrastructure for Computing (SNIC).
AI is funded by the Generalitat Valenciana (Spain) under the grant number CIDEGENT/2020/21. AI also acknowledges the financial support from the MCIN with funding from the European Union NextGenerationEU and Generalitat Valenciana in the call Programa de Planes Complementarios de I+D+i (PRTR 2022), reference ASFAE$/2022/015$

\printbibliography

\end{document}